# Metasurface-generated large and arbitrary analog convolution kernels for accelerated machine vision

Ruiqi Liang[1,2], Shuai Wang[1,3], Yiying Dong[1], Liu Li[1], Ying Kuang[1], Bohan Zhang[1], Yuanmu Yang[1]*

[1]State Key Laboratory of Precision Measurement Technology and Instruments, Department of Precision Instrument, Tsinghua University, Beijing 100084, China
[2]Weiyang College, Tsinghua University, Beijing 100084, China
[3]Qingdao Innovation and Development Center, Harbin Engineering University, Qingdao 26600, China
*ymyang@tsinghua.edu.cn

**In the rapidly evolving field of artificial intelligence, convolutional neural networks are essential for tackling complex challenges such as machine vision and medical diagnosis. Recently, to address the challenges in processing speed and power consumption of conventional digital convolution operations, many optical components have been suggested to replace the digital convolution layer in the neural network, accelerating various machine vision tasks. Nonetheless, the analog nature of the optical convolution kernel has not been fully explored. Here, we develop a spatial frequency domain training method to create arbitrarily shaped analog convolution kernels using an optical metasurface as the convolution layer, with its receptive field largely surpassing digital convolution kernels. By employing spatial multiplexing, the multiple parallel convolution kernels with both positive and negative weights are generated under the incoherent illumination condition. We experimentally demonstrate a 98.59% classification accuracy on the MNIST dataset, with simulations showing 92.63% and 68.67% accuracy on the Fashion-MNIST and CIFAR-10 datasets with additional digital layers. This work underscores the unique advantage of analog optical convolution, offering a promising avenue to accelerate machine vision tasks, especially in edge devices.**

## Introduction

Neural networks, with their ability to be trained from data and make intelligent decisions, are instrumental in solving complex problems such as machine vision[1,2], autonomous driving[3,4], and large language model[5], driving advancements in artificial intelligence. However, the high throughput data processing imposes major challenges in processing speed and power consumption[6], especially in edge devices[7]. Therefore, there has been a growing interest in developing alternative methods to accelerate processing speed and reduce power consumption for neural networks.

Optics, known for its high speed, low power consumption, large bandwidth, and multiple degrees of freedom, presents a potential solution to accelerate neural networks[8,9], which has been demonstrated in both on-chip[10-16] and free-space diffactive[17-24] configurations. Specifically, free-space diffractive neural networks hold higher computation density and are compatible with various machine vision tasks. However, despite some recent progress[25-30], all-optical diffractive neural networks generally lack nonlinearity and reconfigurability, which prevents their applications in handling complex classification or segmentation tasks.

The optoelectronic hybrid neural network combines the flexible digital layers with the optical neural networks, therefore can achieve better performance in various machine vision tasks. Based on the theory that convolution in the spatial domain is equivalent to multiplication in the spatial frequency domain, parallel optical convolution systems[31-34] were designed to replace digital convolution, which comprises a majority of floating-point operations (FLOPs). However, the bulky 4-*f* system and the need for coherent illumination present significant challenges in practical applications. Recently, compact optical neural networks have been devised based on the optical mask[35-37] with amplitude modulation, which greatly shrinks the size and volume of the optical system. However, amplitude modulation may result in significant energy loss compared with phase modulation.

Metasurface provides a compact platform for phase modulation[38] and offers the ability to utilize multiple degrees of freedom of light, such as polarization[39,40], wavelength[41,42], and angle of incidence[43] to enrich modulation dimensions for optical neural networks. In some implementations, the metasurface was used to serve as a single convolution kernel to perform tasks like edge detection[44,45] and classification[46,47]. Alternatively, multi-channel optical digital convolution kernels based on metasurface[48-51] were achieved by polarization and spatial multiplexing, resulting in higher accuracy in classification tasks. Nevertheless, in most implementations, optical convolution kernels were designed to mimic conventional digital convolution kernels (Fig. 1a) with a finite kernel size, up to 7×7 (Fig. 1b), despite that in principle, the size of an optical convolution kernel can be made very large. The limitation is that it becomes progressively more difficult to train a neural network with an increasing kernel size[52].

In this work, we propose and experimentally realize an analog convolutional optoelectronic hybrid neural network (ACNN) with metasurface-generated large and arbitrary analog convolution kernels (Fig. 1c). We develop a frequency domain training method to build convolution kernels with arbitrary shapes and large receptive fields. To realize the convolution kernels with both positive and negative weights, synthetic convolution kernels are designed via spatial multiplexing. With the multi-channel convolution layer connected to a digital backend, we achieve 98.59% classification accuracy for the MNIST dataset in the experiment. Compared to a digital neural network with a similar architecture, there is a 97% reduction in FLOPs, leading to a significant speedup in classification tasks. Further simulation shows a 92.63% classification accuracy for the Fashion-MNIST dataset and 68.67% classification accuracy for the CIFAR-10 datasets with additional digital layers, respectively.

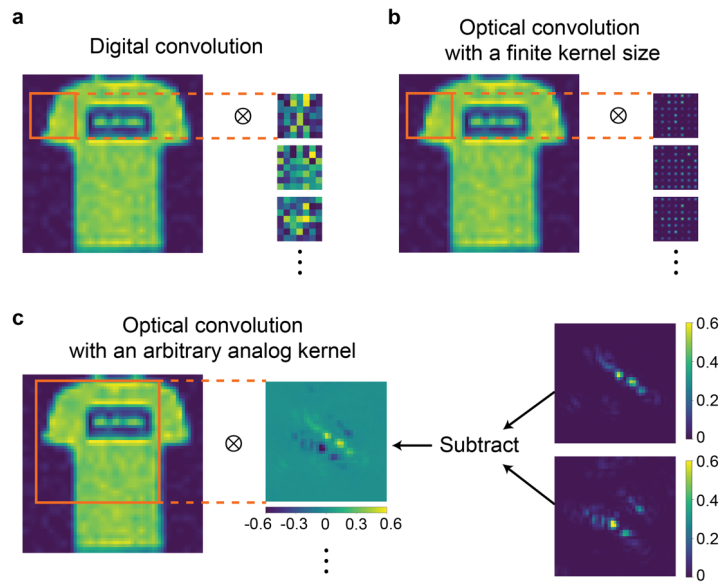

**Figure 1 | Comparison among three different types of convolution kernels. a,** Digital convolution using multiple convolution kernels with a finite kernel size. **b,** Optical convolution using multiple convolution kernels with a finite kernel size. **c,** Optical convolution using multiple convolution kernels with a large and

arbitrary kernel size. The kernel weights with both positive and negative values are realized by subtracting two positive kernels.

## Results

**System configuration of the ACNN**

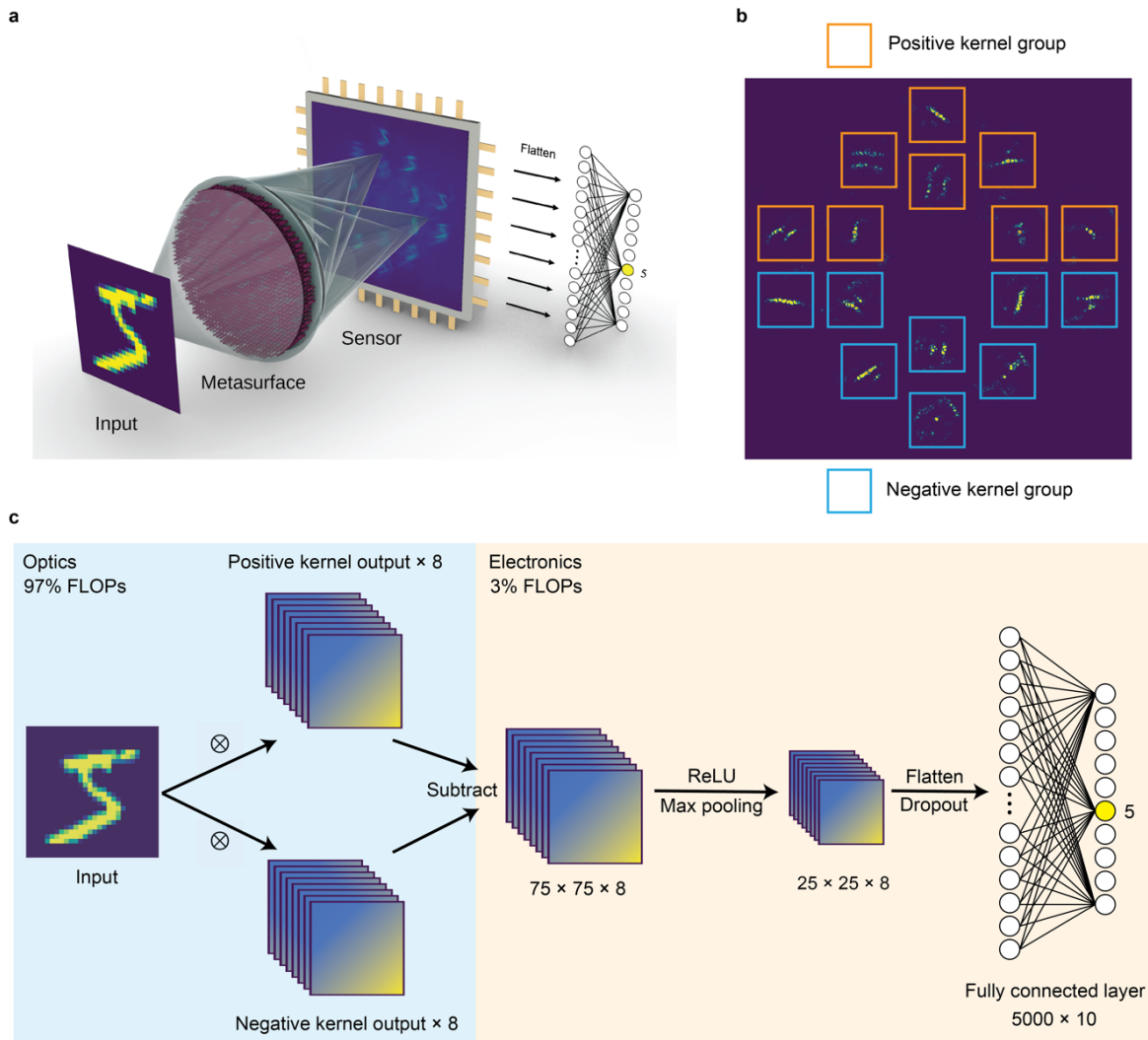

**Figure 2 | System configuration and network structure of the ACNN. a,** Schematic of the system using a metasurface to perform multi-kernel convolution of the input image and project all features to the sensor. The features are then fed into a digital backend for classification. **b,** Kernels (PSF) distribution of the ACNN. Kernels highlighted in orange and blue frames have positive and negative weights, respectively. **c,** Network structure with convolution operations done in optics, which takes up most of the FLOPs. A digital backend, including ReLU activation, max pooling, flattening, dropout, and fully connected layer, performs classification after the optical convolution.

The ACNN consists of a metasurface-based convolution layer and a digital backend, including a fully connected layer and other digital operations (Fig. 2a). Conventional convolution operation in electronics involves a small $N×N$ kernel multiplying and sliding over the target images. An optical element, such as a lens, can be used to mimic a convolution kernel since the relationship between object and image is also a convolution operation, described by,

$$I_i(x, y) = I_o(x, y) \otimes PSF(x, y), \tag{1}$$

where $(x, y)$ is the coordinate of the image plane, $I_i(x, y)$ is the image intensity and $I_o(x, y)$ is the object's ideal image intensity predicted by geometrical optics. $PSF(x, y)$ is the point spread function (PSF) of the optical system, which plays the same role as a convolution kernel (Supplementary Section 1).

Even for a simple classification task on the MNIST dataset, it is necessary to have multiple kernels in the convolution layer to extract various features of an image. This challenge can be addressed by leveraging the unique capability of a metasurface to generate multiple different PSFs (convolution kernels) via spatial multiplexing (Supplementary Section 2). Another issue is that under incoherent illumination, the PSF of an optical system is described by the intensity distribution with positive values. To realize negative kernel weights using the PSF, we divide all parallel kernels into two groups: one group with positive kernel weights and the other with negative kernel weights (Fig. 2b). Subsequently, the synthetic convolution kernels, which incorporate both positive and negative weights, are derived by subtracting one group from the other. In the following step, we perform ReLU activation, max pooling, flattening, and dropout operations on the 8 feature maps. The resulting output is then fed into a fully connected layer for classification (Fig. 2c). For more complicated classification tasks on large-size datasets such as Fashion-MNIST and CIFAR-10, additional convolution and fully connected layers can be added to achieve better performance.

**Training of arbitrary shape convolution kernels in the spatial frequency domain**

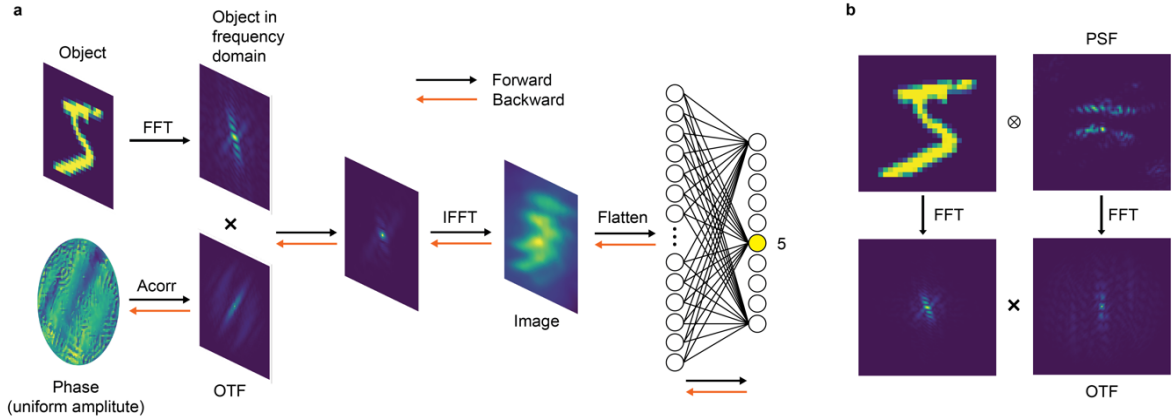

**Figure 3 | Training of the ACNN. a**, Training in the spatial frequency domain involves performing a Fast Fourier Transform (FFT) on the object to obtain its frequency domain distribution, which is multiplied by the OTF derived from the Normalized Autocorrelation Function (Acorr) of the pupil phase distribution. An Inverse Fast Fourier Transform (IFFT) is then performed to generate the final image, which is flattened and fed into the digital backend for classification. The black and orange arrows represent the forward model and the backpropagation-based end-to-end training process, respectively. **b**, An arbitrarily shaped analog kernel with a large receptive field in the spatial domain is equivalent to a large tunable area with more trainable parameters in the spatial frequency domain.

Conventionally, training a convolution neural network involves applying the backpropagation method to find the most suitable convolution kernels for a specific task[49]. With the target convolution kernels (PSFs), iterative algorithms can be used to identify the phase distribution of the metasurface[50], which benefits from high transmittance in phase-only metasurface. However, there are two issues associated with this process. First of all, the traditional backpropagation method only works well with finite-size convolution kernels. When training a convolution neural network with large kernels, which is beneficial for many classification and segmentation tasks[52] and can be easily implemented with optics, it requires complex processes like re-parameterization[50] and may result in divergence. Moreover, even if the kernels are well-trained, there is an additional challenge in realizing the target PSFs via holographic projection.

To tackle the abovementioned challenges, we develop an end-to-end frequency domain training method to build convolution kernels with arbitrary shapes and large receptive fields. The spatial frequency domain, with its intrinsic global receptive field, can be implemented to train larger kernels without the need for complex processes. Additionally, based on the physical propagation model, the training method in the spatial frequency domain provides an end-to-end approach to obtain the phase without requiring holographic reconstruction.

Equation (1) can be described in the spatial frequency domain as,

$$G_i(f_x, f_y) = G_o(f_x, f_y) OTF(f_x, f_y), \qquad (2)$$

where $(f_x, f_y)$ is the spatial frequency coordinate of $(x, y)$, $G_i(f_x, f_y)$ is the Fourier transform of the image intensity, $G_o(f_x, f_y)$ is the Fourier transform of object's ideal image intensity, and $OTF(f_x, f_y)$ is the optical transfer function (OTF) which is also the Fourier transform of $PSF(x, y)$.

We use pupil function (P) to relate the phase of the metasurface with OTF. P is defined as,

$$P(x_m, y_m) = A_p e^{i\theta_p(x_m, y_m)}, \qquad (3)$$

where $A_p$ is the amplitude, which is treated as uniform, and $\theta_p(x_m, y_m)$ is the phase distribution. The OTF relates to P via the autocorrelation function as,

$$OTF(f_x, f_y) = \iint_{-\infty}^{\infty} P(x_m + \lambda z_i f_x/2, y_m + \lambda z_i f_y/2) P^*(x_m - \lambda z_i f_x/2, y_m - \lambda z_i f_y/2) dx_m dy_m, \qquad (4)$$

where $(x_m, y_m)$ is the coordinate of the metasurface, $z_i$ is the distance between the metasurface and the sensor, and $\lambda$ is the working wavelength, which is 632 nm (Supplementary Section 3).

Equations (1)-(4) describe the forward propagation model in the frequency domain, as illustrated in Fig. 3a. The orange arrow in Fig. 3a shows the backpropagation process to train the phase distribution of the metasurface. Due to the global receptive field of the spatial frequency domain and the end-to-end training based on physical propagation, this method allows us to obtain optical convolution kernels with large receptive fields and arbitrary shapes in the spatial domain. As shown in Fig. 3b, the large tunable area in the frequency domain enables the optical system to have more trainable parameters (Supplementary Section 4).

# Experimental results

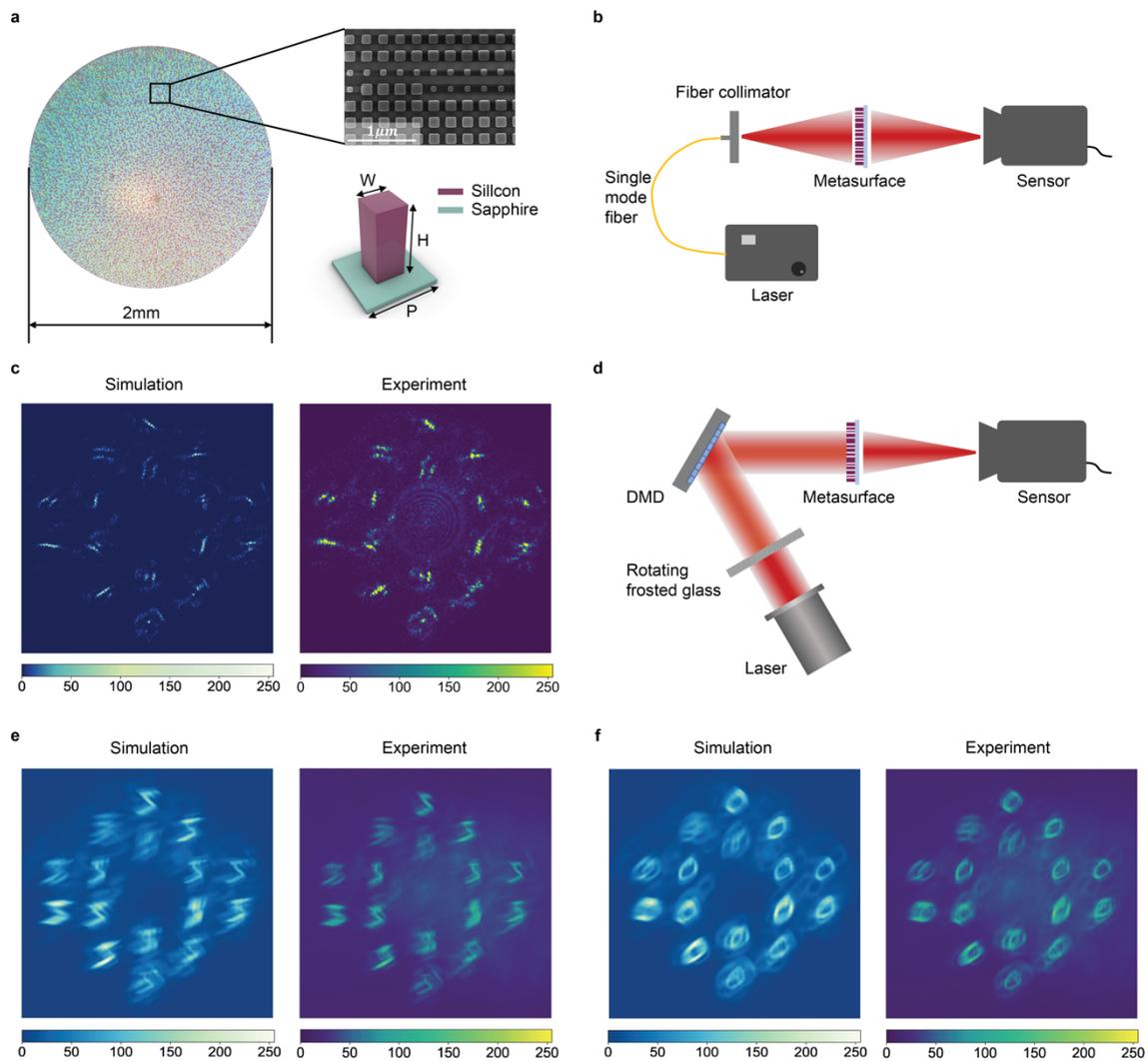

**Figure 4 | Experiment setup and characterization of the metasurface. a,** Optical microscopy (left) and scanning electron microscopy (upper right) images of the metasurface, respectively. The unit cell of the metasurface (lower right) is composed of silicon nanopillars on a sapphire substrate, with height $H$ = 300 nm and period $P$ = 250 nm. The width $W$ varies between 70 nm and 160 nm. **b,** Experiment setup for the PSF characterization. **c,** Simulated (left) and measured PSF (right), respectively. **d,** Experiment setup for the ACNN-based image classification. **e,** Simulated (left) and measured (right) images of input '5' after convolution, respectively. **f,** Simulated (left) and measured (right) images of input '0' after convolution, respectively.

The metasurface with a diameter of 2 mm is fabricated using standard electron-beam lithography and reactive-ion etching. It consists of a sapphire ($Al_2O_3$) substrate and 300-nm-thick crystalline silicon nano-pillars with a period of 260 nm. The phase retardation and transmissivity of 632 nm light through the unit cell are obtained from numerical simulation by sweeping the width $W$ of the crystalline silicon nano-pillar from 70 nm to 160 nm (Supplementary Section 5). The optical microscopy image and scanning electron microscopy image of the fabricated metasurface are shown in Fig. 4a.

The PSF of the metasurface is measured using the setup as illustrated in Fig. 4b, which matches well with the simulated PSF, as shown in Fig. 4c. For the classification of the MNIST dataset, the experimental setup is depicted in Fig. 4d. To generate incoherent illumination, a rotating frosted glass is placed in front of the laser source. The input images are subsequently loaded onto the Digital Micromirror Device (DMD), convoluted with the metasurface, and captured by a CMOS image sensor. Figures 4e-f show two experimentally obtained images, which match well to the simulation, with some background noise mainly caused by the zeroth-order diffraction of the metasurface. Note to mitigate the effect of the background noise, we specifically design the PSF distribution to avoid the central region of the image sensor.

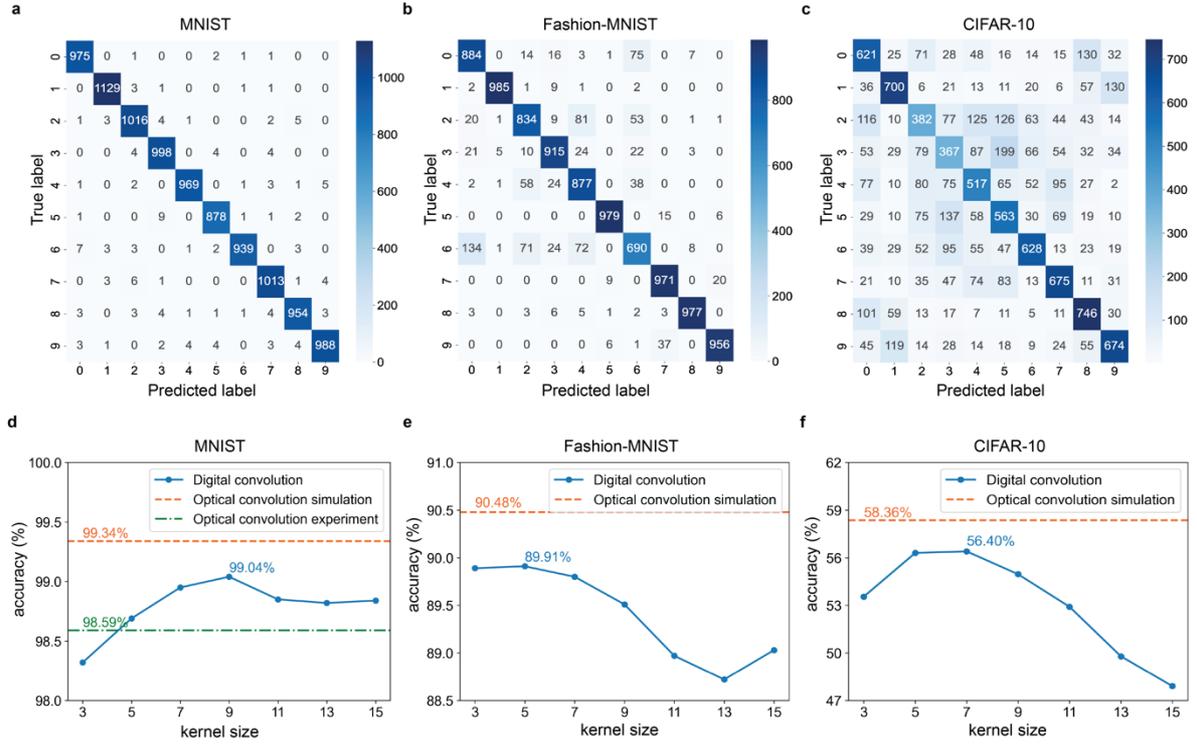

**Figure 5 | Classification tasks using ACNN. a,** Experimental confusion matrix using validation images of the MNIST dataset. **b,** Simulated confusion matrix using validation images of the Fashion-MNIST dataset. **c,** Simulated confusion matrix using validation images of the CIFAR-10 dataset. **d**, Comparison of classification accuracy as a function of convolution kernel size using the MNIST dataset. **e**, Comparison of classification accuracy as a function of convolution kernel size using the Fashion-MNIST dataset. **f**, Comparison of classification accuracy as a function of convolution kernel size using the CIFAR-10 dataset.

To demonstrate the performance of the ACNN, we applied it to the classification task. The MNIST dataset contains 60,000 handwritten digit images as the training dataset and 10,000 handwritten digit images as the validation dataset. In the simulation, the ACNN achieves 99.34% classification accuracy on the validation dataset. Experimentally, it achieves 98.59% classification accuracy on the validation dataset after the re-training process of the digital backend. The experimental confusion matrix for the validation dataset is shown in Fig. 5a. The slight decrease in classification accuracy in the experiment is attributed to deviations in realized optical kernels.

To benchmark the speedup in the MNIST classification task using the ACNN, consider a fully digital neural network with an identical configuration, and substitute the optical kernels with 8 digital kernels of size 3×3. In this scenario, the ACNN achieves a 78% reduction in FLOPs. If the digital kernels have a size of 9×9, the ACNN can further achieve a 97% reduction in FLOPs (Supplementary Section 6).

To demonstrate the versatility of ACNN, we perform simulations on more complex Fashion-MNIST and CIFAR-10 datasets with an identical network structure. The Fashion-MNIST dataset consists of 60,000 training images and 10,000 validation images across 10 different categories of clothing items. The ACNN achieves 90.61% classification accuracy on the validation images with the confusion matrix shown in Fig. 5b. The CIFAR-10 dataset consists of 60,000 color images (converted to grayscale in our simulation) in 10 different classes, with 50,000 training images and 10,000 validation images. The ACNN achieves 58.36% accuracy on the validation images with the confusion matrix shown in Fig. 5c. To achieve higher classification accuracy, an additional digital convolution layer and two fully connected layers can be added into the neural network, which improves its classification accuracy to 92.63% and 68.67% in the Fashion-MNIST and CIFAR-10 datasets, respectively (Supplementary Section 7).

To verify the advantage of having well-trained large and arbitrary optical convolution kernels, we compare the classification accuracy of the ACNN to a digital convolution neural network with an identical network structure but varying kernel size. As is shown in Figs. 5d-f, for a digital convolution neural network, its classification accuracy initially increases as the kernel size increases, yet such a trend converges or reverses up to a certain point since further increasing the kernel size can cause optimization problems[52]. In comparison, the large and arbitrary optical convolution kernels result in generally higher classification accuracies, while the frequency domain training method avoids optimization problems in training large kernels in the spatial domain. To further test the robustness of the ACNN against quantization error, we perform classification on the MNIST dataset with differently quantified image data format and show that the ACNN can achieve a high classification accuracy of 96.58% even with a 2-bit image (Supplementary Section 8).

## Conclusion

To summarize, we demonstrate a metasurface-based ACNN with an optical convolution layer and a digital backend. A frequency domain training method is developed to create convolution kernels with large receptive fields and arbitrary shapes while avoiding the need for complex reparameterization and artificial reconstruction[50,53]. A 98.59% classification accuracy is demonstrated on the MNIST dataset, with additional simulation showing the advantage of having well-trained large and arbitrary optical convolution kernels.

With a single convolution layer replaced by optics, the speedup of many machine vision tasks, which are based on large-scale networks, may be limited. Nevertheless, it can be adapted for small, lightweight networks suitable for edge devices, utilizing optical acceleration to offload a large portion of the convolution operations[54]. It is possible to accommodate more convolution kernels to fit the need for alternative tasks by decreasing the size of each individual kernel. Furthermore, with the unique advantage of large kernels in many downstream tasks, the application of ACNN may be extended beyond classification to other tasks, such as detection and segmentation[52]. Lastly, given the metasurface's ability to modulate multi-dimensional light fields, ACNN has the potential to be applied in unique tasks such as polarization-based imaging through scattering media[55,56], spectral information-based object detection[57,58], and depth sensing[59].

# References

1. He, K.; Zhang, X.; Ren, S.; Sun, J. Deep Residual Learning for Image Recognition. *Proceedings of the IEEE Conference on Computer Vision and Pattern Recognition*; IEEE, 2016; pp 770−778.

2. Long, J.; Shelhamer, E.; Darrell, T. Fully Convolutional Networks for Semantic Segmentation. *Proceedings of the IEEE Conference on Computer Vision and Pattern Recognition*; IEEE, 2015; pp 3431−3440.

3. Yurtsever, E.; Lambert, J.; Carballo, A.; Takeda, K. A Survey of Autonomous Driving: Common Practices and Emerging Technologies. *IEEE Access* **2020**, *8*, 58443-58469.

4. Liu, L.; Lu, S.; Zhong, R.; Wu, B.; Yao, Y.; Zhang, Q.; Shi, W. Computing Systems for Autonomous Driving: State of the Art and Challenges. *IEEE Internet of Things Journal* **2020**, *8*, 6469−6486.

5. Devlin, J.; Chang, M.-W.; Lee, K.; Toutanova, K. BERT: Pre-training of Deep Bidirectional Transformers for Language Understanding. *Proceedings of the 2019 Conference of the North American Chapter of the Association for Computational Linguistics: Human Language Technologies*; Association for Computational Linguistics: Stroudsburg, 2019; pp 4171−4186.

6. Alzubaidi, L.; Zhang, J.; Humaidi, A. J.; Al-Dujaili, A.; Duan, Y.; Al-Shamma, O.; Santamaría, J.; Fadhel, M. A.; Al-Amidie, M.; Farhan, L. Review of Deep Learning: Concepts, CNN Architectures, Challenges, Applications, Future Directions. *J. Big Data* **2021**, *8*, 1-74.

7. Véstias, M. P. A Survey of Convolutional Neural Networks on Edge with Reconfigurable Computing. *Algorithms* **2019**, *12*, 154.

8. Shastri, B. J.; Tait, A. N.; Ferreira de Lima, T.; Pernice, W. H. P.; Bhaskaran, H.; Wright, C. D.; Prucnal, P. R. Photonics for Artificial Intelligence and Neuromorphic Computing. *Nat. Photonics* **2021**, *15*, 102-114.

9. Wetzstein, G.; Ozcan, A.; Gigan, S.; Fan, S.; Englund, D.; Soljačić, M.; Denz, C.; Miller, D. A. B.; Psaltis, D. Inference in Artificial Intelligence with Deep Optics and Photonics. *Nature* **2020**, *588*, 39-47.

10. Shen, Y.; Harris, N. C.; Skirlo, S.; Prabhu, M.; Baehr-Jones, T.; Hochberg, M.; Sun, X.; et al. Deep Learning with Coherent Nanophotonic Circuits. *Nat. Photonics* **2017**, *11*, 441-446.

11. Feldmann, J.; Youngblood, N.; Karpov, M.; Gehring, H.; Li, X.; Stappers, M.; Le Gallo, M.; et al. Parallel Convolutional Processing Using an Integrated Photonic Tensor Core. *Nature* **2021**, *589*, 52-58.

12. Xu, X.; Tan, M.; Corcoran, B.; Wu, J.; Boes, A.; Nguyen, T. G.; Chu, S. T.; et al. 11 TOPS Photonic Convolutional Accelerator for Optical Neural Networks. *Nature* **2021**, *589*, 44-51.

13. Ashtiani, F.; Geers, A. J.; Aflatouni, F. An On-Chip Photonic Deep Neural Network for Image Classification. *Nature* **2022**, *606*, 501-506.

14. Chen, Y.; Nazhamaiti, M.; Xu, H.; Meng, Y.; Zhou, T.; Li, G.; Fan, J.; et al. All-Analog Photoelectronic Chip for High-Speed Vision Tasks. *Nature* **2023**, *623*, 48-57.

15. Xu, Z.; Zhou, T.; Ma, M.; Deng, C.; Dai, Q.; Fang, L. Large-Scale Photonic Chiplet Taichi Empowers 160-TOPS/W Artificial General Intelligence. *Science* **2024**, *384*, 202-209.

16. Liao, K.; Dai, T.; Yan, Q.; Hu, X.; Gong, Q. Integrated Photonic Neural Networks: Opportunities and Challenges. *ACS Photonics* **2023**, *10*, 2001−2010.

17. Lin, X.; Rivenson, Y.; Yardimci, N. T.; Veli, M.; Luo, Y.; Jarrahi, M.; Ozcan, A. All-Optical Machine Learning Using Diffractive Deep Neural Networks. *Science* **2018**, *361*, 1004-1008.

18. Zhou, T.; Lin, X.; Wu, J.; Chen, Y.; Xie, H.; Li, Y.; Fan, J.; Wu, H.; Fang, L.; Dai, Q. Large-Scale Neuromorphic Optoelectronic Computing with a Reconfigurable Diffractive Processing Unit. *Nat. Photonics* **2021**, *15*, 367-373.

19  Chen, R.; Li, Y.; Lou, M.; Fan, J.; Tang, Y.; Sensale-Rodriguez, B.; Yu, C.; Gao, W. Physics-Aware Machine Learning and Adversarial Attack in Complex-Valued Reconfigurable Diffractive All-Optical Neural Network. *Laser Photonics Rev.* **2022**, *16*, 2200348.

20  Luo, X.; Hu, Y.; Ou, X.; Li, X.; Lai, J.; Liu, N.; Cheng, X.; Pan, A.; Duan, H. Metasurface-Enabled On-Chip Multiplexed Diffractive Neural Networks in the Visible. *Light Sci. Appl.* **2022**, *11*, 158.

21  Liu, C.; Ma, Q.; Luo, Z. J.; Hong, Q. R.; Xiao, Q.; Zhang, H. C.; Miao, L.; et al. A Programmable Diffractive Deep Neural Network Based on a Digital-Coding Metasurface Array. *Nat. Electron.* **2022**, *5*, 113-122.

22  Rahman, M. S. S.; Yang, X.; Li, J.; Bai, B.; Ozcan, A. Universal Linear Intensity Transformations Using Spatially Incoherent Diffractive Processors. *Light Sci. Appl.* **2023**, *12*, 195.

23  Integration of Programmable Diffraction with Digital Neural Networks. *ACS Photonics* **2024**, *11*, 2906−2922.

24  Zuo, Y.; Li, B.; Zhao, Y.; Jiang, Y.; Chen, Y.-C.; Chen, P.; Jo, G.-B.; Liu, J.; Du, S. All-Optical Neural Network with Nonlinear Activation Functions. *Optica* **2019**, *6*, 1132-1137.

25  Ryou, A.; Whitehead, J.; Zhelyeznyakov, M.; Anderson, P.; Keskin, C.; Bajcsy, M.; Majumdar, A. Free-Space Optical Neural Network Based on Thermal Atomic Nonlinearity. *Photonics Res.* **2021**, *9*, B128-B134.

26  Yildirim, M.; Dinc, N. U.; Oguz, I.; Psaltis, D.; Moser, C. Nonlinear Processing with Linear Optics. *Nat. Photonics* **2024**, *1*, 1-7.

27  Wanjura, C. C.; Marquardt, F. Fully Nonlinear Neuromorphic Computing with Linear Wave Scattering. *Nat. Phys.* **2024**, *20*, 1434–1440.

28  Li, Y.; Li, J.; Ozcan, A. Nonlinear Encoding in Diffractive Information Processing Using Linear Optical Materials. *Light Sci. Appl.* **2024**, *13*, 173.

29  Huang, Z.; Shi, W.; Wu, S.; Wang, Y.; Yang, S.; Chen, H. Pre-sensor Computing with Compact Multilayer Optical Neural Network. *Sci. Adv.* **2024**, *10*, eado8516.

30  Wang, T.; Sohoni, M. M.; Wright, L. G.; Stein, M. M.; Ma, S.-Y.; Onodera, T.; Anderson, M. G.; McMahon, P. L. Image Sensing with Multilayer Nonlinear Optical Neural Networks. *Nat. Photonics* **2023**, *17*, 408−415.

31  Chang, J.; Sitzmann, V.; Dun, X.; Heidrich, W.; Wetzstein, G. Hybrid Optical-Electronic Convolutional Neural Networks with Optimized Diffractive Optics for Image Classification. *Sci. Rep.* **2018**, *8*, 1-10.

32  Colburn, S.; Chu, Y.; Shilzerman, E.; Majumdar, A. Optical Frontend for a Convolutional Neural Network. *Appl. Opt.* **2019**, *58*, 3179-3186.

33  Miscuglio, M.; Hu, Z.; Li, S.; George, J. K.; Capanna, R.; Dalir, H.; Bardet, P. M.; Gupta, P.; Sorger, V. J. Massively Parallel Amplitude-Only Fourier Neural Network. *Optica* **2020**, *7*, 1812-1819.

34  Bernstein, L.; Sludds, A.; Panuski, C.; Trajtenberg-Mills, S.; Hamerly, R.; Englund, D. Single-Shot Optical Neural Network. *Sci. Adv.* **2023**, *9*, eadg7904.

35  Chen, H. G.; Jayasuriya, S.; Yang, J.; Stephen, J.; Sivaramakrishnan, S.; Veeraraghavan, A.; Molnar, A. ASP Vision: Optically Computing the First Layer of Convolutional Neural Networks Using Angle Sensitive Pixels. *Proceedings of the IEEE Conference on Computer Vision and Pattern Recognition*; IEEE, 2016; pp 903−912.

36  Pad, P.; Narduzzi, S.; Kundig, C.; Turetken, E.; Bigdeli, S. A.; Dunbar, L. A. Efficient Neural Vision Systems Based on Convolutional Image Acquisition. *Proceedings of the IEEE/CVF Conference on Computer Vision and Pattern Recognition*; IEEE, 2020; pp 12285−12294.

37  Shi, W.; Huang, Z.; Huang, H.; Hu, C.; Chen, M.; Yang, S.; Chen, H. LOEN: Lensless Opto-Electronic Neural Network Empowered Machine Vision. *Light Sci. Appl.* **2022**, *11*, 121.

38  Chen, H.-T.; Taylor, A. J.; Yu, N. A Review of Metasurfaces: Physics and Applications. *Rep. Prog. Phys.* **2016**, *79*, 076401.

39  Hu, Y.; Wang, X.; Luo, X.; Ou, X.; Li, L.; Chen, Y.; Yang, P.; Wang, S.; Duan, H. All-Dielectric Metasurfaces for Polarization Manipulation: Principles and Emerging Applications. *Nanophotonics* **2020**, *9*, 3755-3780.

40  Wang, S.; Wen, S.; Deng, Z.-L.; Li, X.; Yang, Y. Metasurface-Based Solid Poincaré Sphere Polarizer. *Phys. Rev. Lett.* **2023**, *130*, 123801.

41  Khorasaninejad, M.; Chen, W. T.; Devlin, R. C.; Oh, J.; Zhu, A. Y.; Capasso, F. Metalenses at Visible Wavelengths: Diffraction-Limited Focusing and Subwavelength Resolution Imaging. *Science* **2016**, *352*, 1190-1194.

42  Shi, Z.; Khorasaninejad, M.; Huang, Y. W.; Roques-Carmes, C.; Zhu, A. Y.; Chen, W. T.; Capasso, F.; et al. Single-Layer Metasurface with Controllable Multiwavelength Functions. *Nano Lett.* **2018**, *18*, 2420-2427.

43  Levanon, N.; Indukuri, S. C.; Frydendahl, C.; Bar-David, J.; Han, Z.; Mazurski, N.; Levy, U. Angular Transmission Response of In-Plane Symmetry-Breaking Quasi-BIC All-Dielectric Metasurfaces. *ACS Photonics* **2022**, *9*, 3642-3648.

44  Fu, W.; Zhao, D.; Li, Z.; Liu, S.; Tian, C.; Huang, K. Ultracompact Meta-Imagers for Arbitrary All-Optical Convolution. *Light Sci. Appl.* **2022**, *11*, 62.

45  Wang, S.; Li, L.; Wen, S.; Liang, R.; Liu, Y.; Zhao, F.; Yang, Y. Metalens for Accelerated Optoelectronic Edge Detection under Ambient Illumination. *Nano Lett.* **2023**, *24*, 356-361.

46  Luo, X.; Hu, Y.; Ou, X.; Li, X.; Lai, J.; Liu, N.; Cheng, X.; Pan, A.; Duan, H. Metasurface-enabled on-chip multiplexed diffractive neural networks in the visible. *Light Sci. Appl.* **2022**, *11*, 158.

47  Huang, L.; Tanguy, Q. A. A.; Fröch, J. E.; Mukherjee, S.; Böhringer, K. F.; Majumdar, A. Photonic Advantage of Optical Encoders. *Nanophotonics* **2024**, *13*, 1191-1196.

48  Zheng, H.; Liu, Q.; Zhou, Y.; Kravchenko, I. I.; Huo, Y.; Valentine, J. Meta-Optic Accelerators for Object Classifiers. *Sci. Adv.* **2022**, *8*, eabo6410.

49  Zheng, H.; Liu, Q.; Kravchenko, I. I.; Zhang, X.; Huo, Y.; Valentine, J. G. Multichannel Meta-Imagers for Accelerating Machine Vision. *Nat. Nanotechnol.* **2024**, *19*, 471-478.

50  Wei, Kaixuan, Xiao Li, Johannes Froech, Praneeth Chakravarthula, James Whitehead, Ethan Tseng, Arka Majumdar, and Felix Heide. Spatially varying nanophotonic neural networks. arXiv:2308.03407 *[cs.CV]*, 2023, na.

51  Luo, M.; Xu, T.; Xiao, S.; Tsang, H. K.; Shu, C.; Huang, C. Meta-Optics Based Parallel Convolutional Processing for Neural Network Accelerator. *Laser Photonics Rev.* **2024**, 2300984.

52  Ding, X.; Zhang, X.; Han, J.; Ding, G. Scaling Up Your Kernels to 31x31: Revisiting Large Kernel Design in CNNs. *Proceedings of the IEEE/CVF Conference on Computer Vision and Pattern Recognition*; IEEE, 2022; pp 11963−11975.

53  Liu, Q.; Zheng, H.; Swartz, B. T.; Lee, H. H.; Asad, Z.; Kravchenko, I.; Valentine, J. G.; Huo, Y. Digital Modeling on Large Kernel Metamaterial Neural Network. *J. Imaging Sci. Technol.* **2023**, *67*.

54  Liu, Q.; Swartz, B. T.; Kravchenko, I.; Valentine, J. G.; Huo, Y. ExtremeMETA: High-speed Lightweight Image Segmentation Model by Remodeling Multi-channel Metamaterial Imagers. arXiv:2405.17568 *[cs.CV]*, 2024, na.

55  Liu, H.; Wang, F.; Jin, Y.; Ma, X.; Li, S.; Bian, Y.; Situ, G. Learning-Based Real-Time Imaging Through Dynamic Scattering Media. *Light Sci. Appl*. **2024**, *13*, 194.

56  Zaidi, A.; Rubin, N. A.; Meretska, M. L.; Li, L. W.; Dorrah, A. H.; Park, J. S.; Capasso, F. Metasurface-Enabled Single-Shot and Complete Mueller Matrix Imaging. *Nat. Photonics* **2024**, *18*, 704-712.

57  McClung, A.; Samudrala, S.; Torfeh, M.; Mansouree, M.; Arbabi, A. Snapshot Spectral Imaging with Parallel Metasystems. *Sci. Adv*. **2020**, *6*, eabc7646.

58  He, X.; Tang, C.; Liu, X.; Zhang, W.; Sun, K.; Xu, J. Object Detection in Hyperspectral Image via Unified Spectral–Spatial Feature Aggregation. *IEEE Transactions on Geoscience and Remote Sensing* **2023**, *61*, 1−13.

59  Shen, Z.; Zhao, F.; Jin, C.; Wang, S.; Cao, L.; Yang, Y. Monocular Metasurface Camera for Passive Single-Shot 4D Imaging. *Nat. Commun*. **2023**, *14*, 1035.

## Data availability

All relevant data are available in the main text, in the Supporting Information, or from the authors. The source code of the ACNN is available at https://github.com/THUMetaOptics/ACNN.

## Acknowledgment

This work was supported by the National Key Research and Development Program of China (2023YFB2805800), Beijing Municipal Science & Technology Commission, Administrative Commission of Zhongguancun Science Park (No. Z231100006023006), and Beijing Innovation Training Program for College Students (No. 202310003079).

## Author contributions

Y.Y. and R.L. conceived this work. R.L. and S.W. developed the frequency domain training method and designed the metasurface; R.L., Y.D., and B.Z optimized the training method; Y.K. did the Zemax simulation. R.L. and L.L. conducted the experiment; R.L. and Y.Y. wrote the manuscript. Y.Y. supervised the project.

## Competing interests

The authors declare no competing interests.

# Supplementary Information:

# Metasurface-generated large and arbitrary analog convolution kernels for accelerated machine vision

Ruiqi Liang[1,2], Shuai Wang[1], Yiying Dong[1], Liu Li[1], Ying Kuang[1], Bohan Zhang[1], Yuanmu Yang[1]*

[1]State Key Laboratory of Precision Measurement Technology and Instruments, Department of Precision Instrument, Tsinghua University, Beijing 100084, China
[2]Weiyang College, Tsinghua University, Beijing 100084, China
*ymyang@tsinghua.edu.cn

## S1. Fourier Optics-based description of metasurface-based convolution

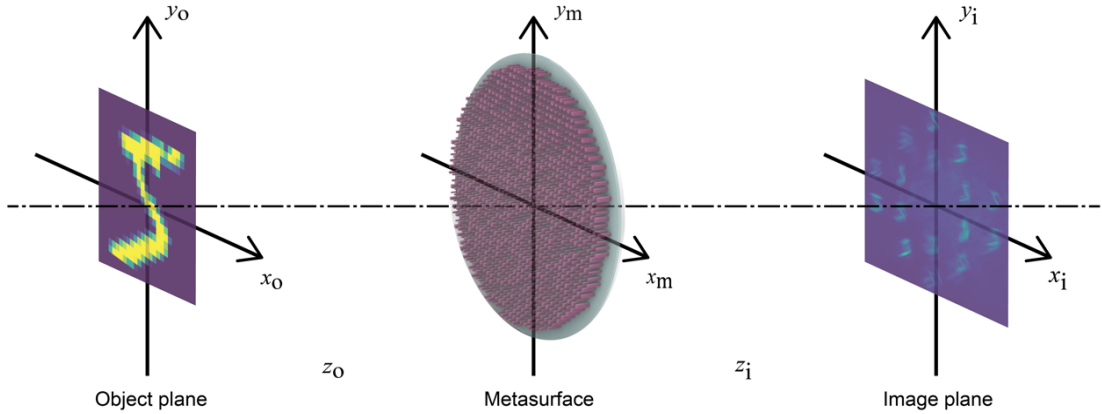

**Figure S1** | Schematic of the propagation model. Images are loaded on the object plane and propagate to the metasurface, which performs multi-channel convolution of the input and projects the convoluted result to the image plane.

As shown in Fig. S1, under the Fresnel approximation, the propagation of a point light source at the center point of the object plane to the metasurface is described as,

$$U_{\mathrm{m}}(x_{\mathrm{m}}, y_{\mathrm{m}}) = \frac{\exp(\mathrm{j}kz_{\mathrm{o}})}{\mathrm{j}\lambda z_{\mathrm{o}}} \exp\left(\mathrm{j}k\frac{x_{\mathrm{m}}^2 + y_{\mathrm{m}}^2}{2z_{\mathrm{o}}}\right), \quad \text{(S1)}$$

where $(x_{\mathrm{m}}, y_{\mathrm{m}})$ is the coordinate of the metasurface, $U_{\mathrm{m}}(x_{\mathrm{m}}, y_{\mathrm{m}})$ is the complex amplitude distribution of light before the metasurface, $k=\frac{2\pi}{\lambda}$ is the wave vector, $\lambda$ is the working wavelength, and $z_{\mathrm{o}}$ is the distance between the object plane and metasurface.

The transmission function of the metasurface is defined as,

$$T_{\mathrm{m}}(x_{\mathrm{m}}, y_{\mathrm{m}}) = P(x_{\mathrm{m}}, y_{\mathrm{m}})\exp\left[-\mathrm{j}\frac{k}{2f}(x_{\mathrm{m}}^2 + y_{\mathrm{m}}^2)\right], \quad \text{(S2)}$$

where $P(x_{\mathrm{m}}, y_{\mathrm{m}})$ is defined as the pupil function of the metasurface, and $f$ is the focal distance of the lens.

The complex amplitude distribution of light through the metasurface is,

$$U'_{\mathrm{m}}(x_{\mathrm{m}}, y_{\mathrm{m}}) = U_{\mathrm{m}}(x_{\mathrm{m}}, y_{\mathrm{m}})T_{\mathrm{m}}(x_{\mathrm{m}}, y_{\mathrm{m}}). \quad \text{(S3)}$$

Under the Fresnel approximation, the propagation of the light from the metasurface to the image plane is described as,

$$U_i(x_i, y_i) = \frac{\exp[jk(z_o + z_i)]}{\lambda^2 z_o z_i} \exp\left(jk\frac{x_i^2 + y_i^2}{2z_i}\right)$$
$$\iint_{-\infty}^{\infty} P(x_m, y_m) \exp\left[j\frac{k}{2}\left(\frac{1}{z_o} + \frac{1}{z_i} - \frac{1}{f}\right)(x_m^2 + y_m^2)\right] \exp\left[-j\frac{k}{z_i}(x_i x_m + y_i y_m)\right] dx_m dy_m, \quad (S4)$$

where $(x_i, y_i)$ is the coordinate of the image plane, $U_i(x_i, y_i)$ is the complex amplitude distribution of the light on the image plane, and $z_i$ is the distance between the metasurface and image plane.

Here, we establish the positions of the object plane, metasurface, and image plane according to the relationship $\frac{1}{z_o} + \frac{1}{z_i} - \frac{1}{f} = 0$. Substitute it into Eq. S4, we obtain,

$$U_i(x_i, y_i) = \frac{\exp[jk(z_o + z_i)]}{\lambda^2 z_o z_i} \exp\left(jk\frac{x_i^2 + y_i^2}{2z_i}\right)$$
$$\iint_{-\infty}^{\infty} P(x_m, y_m) \exp\left[-j\frac{k}{z_i}(x_i x_m + y_i y_m)\right] dx_m dy_m, \quad (S5)$$

where $U_i(x_i, y_i)$ can be regarded as the Fourier transform of the pupil function $P(x_m, y_m)$ of the metasurface.

Under incoherent illumination, the point spread function (PSF) is the modulus squared of $U_i(x_i, y_i)$, which is described as,

$$PSF(x_i, y_i) = |U_i(x_i, y_i)|^2. \quad (S6)$$

Here, the metasurface-based converlution system can been treated as a linear invariant system under the Fresnel approximation. The image of the object under incoherent illumination on the image plane is,

$$I_i(x_i, y_i) = I_o(x_i, y_i) \otimes PSF(x_i, y_i), \quad (S7)$$

where $I_o(x_i, y_i)$ is the ideal image intensity predicted by geometrical optics, which has the enlargement factor as $M = \frac{z_i}{z_o}$, $x_i \approx -Mx_o$, and $y_i \approx -My_o$.

## S2. Spatial multiplexing

We use spatial multiplexing to obtain multi-channel convolution kernels. By controlling the phase distribution of the metasurface, we cast 16 convolution kernels into different areas of the sensor. To achieve the optics convolution in Eq. S13, the phase modulation in the metasurface plane should be,

$$\theta_m(x_m, y_m) = \text{angle}\left\{\sum_{k=1}^{16}\left\{\exp[j\theta_{P_k}(x_m, y_m)]\exp\left[j\frac{2\pi}{f}\left(\frac{x_k}{z_i}x_m + \frac{y_k}{z_i}y_m\right)\right]\right\}\exp\left[-j\frac{2\pi}{\lambda f}(x_m^2 + y_m^2)\right]\right\}, \quad (S8)$$

where $\theta_{P_k}(x_m, y_m)$ is the phase of P for kernel $k$, and $(x_k, y_k)$ is the position of the focal point for kernel $k$ in the image plane.

## S3. Autocorrelation function

To describe Eq. S13 in the spatial frequency domain, the normalized spatial frequency spectra of $I_i(x_i, y_i)$ and $I_o(x_i, y_i)$ are defined as,

$$G_o(f_x, f_y) = \frac{\iint_{-\infty}^{\infty} I_o(x_i, y_i) \exp\left[-j2\pi(f_x x_i + f_y y_i)\right] dx_i dy_i}{\iint_{-\infty}^{\infty} I_o(x_i, y_i) dx_i dy_i}, \quad (S9)$$

$$G_i(f_x, f_y) = \frac{\iint_{-\infty}^{\infty} I_i(x_i, y_i) \exp\left[-j2\pi\left(f_x x_i + f_y y_i\right)\right] dx_i dy_i}{\iint_{-\infty}^{\infty} I_i(x_i, y_i) dx_i dy_i}. \tag{S10}$$

The optical transfer function (OTF) is defined as,

$$OTF(f_x, f_y) = \frac{\iint_{-\infty}^{\infty} |h(x_i, y_i)|^2 \exp\left[-j2\pi\left(f_x x_i + f_y y_i\right)\right] dx_i dy_i}{\iint_{-\infty}^{\infty} |h(x_i, y_i)|^2 dx_i dy_i}. \tag{S11}$$

According to the convolution theorem, Eq. S13 in the spatial frequency domain is described as,

$$G_i(f_x, f_y) = G_o(f_x, f_y) OTF(f_x, f_y). \tag{S12}$$

According to Rayleigh's theorem, the relationship between $OTF(f_x, f_y)$ and $P(x_m, y_m)$ is the normalized autocorrelation function (Acorr) as,

$$OTF(f_x, f_y) = \frac{\iint_{-\infty}^{\infty} P(x_m + \lambda z_i f_x/2, y_m + \lambda z_i f_y/2) P^*(x_m - \lambda z_i f_x/2, y_m - \lambda z_i f_y/2) dx_m dy_m}{\iint_{-\infty}^{\infty} |P(x_m, y_m)|^2 dx_m dy_m}. \tag{S13}$$

## S4. Trainable parameters

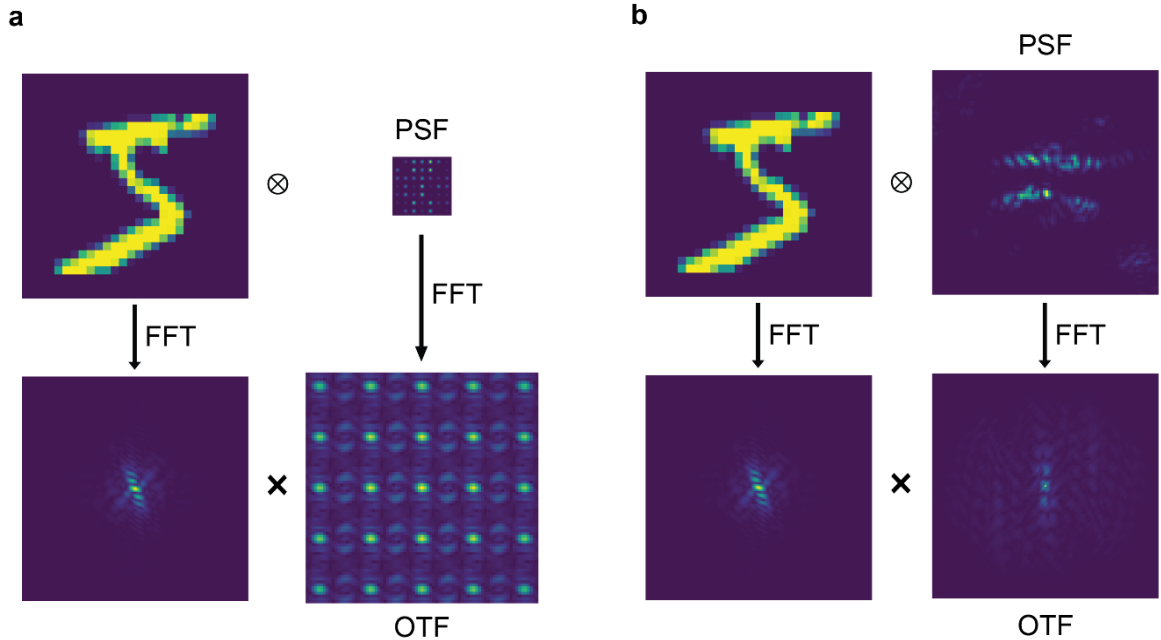

**Figure S2 | a**, A kernel with finite size in the spatial domain is equivalent to a fixed number of points in a fixed pattern in the spatial frequency domain. **b**, An arbitrarily shaped analog kernel with a large receptive field in the spatial domain is equivalent to a large tunable area with more trainable parameters in the spatial frequency domain.

For the method of optical digital convolution, there are several tunable points in the spatial domain, with fixed points in a fixed pattern tunable in the spatial frequency domain, as shown in Fig. S2a. The trainable parameters of a convolution kernel are $N^2$, where $N$ is the kernel size, typically 3, 5, or 7. For a kernel size of 7, the number of trainable parameters of a convolution kernel is 49.

In our method, we train the phase distribution of pupil function end-to-end in the spatial frequency domain. The sampling size of phase distribution is 84×84, resulting in 7056 parameters trainable for one kernel. The large number of trainable parameters allows the kernels to have arbitrary shapes with more details and large receptive fields. In the spatial frequency domain, there is a large tunable area of OTF, as shown in Fig. S2b.

## S5. Design of metasurface unit cell

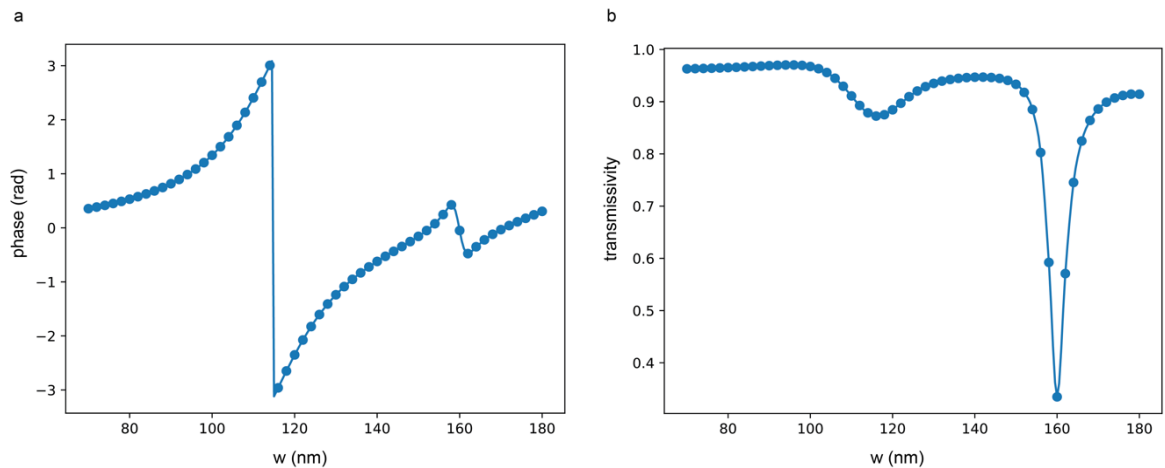

**Figure S3 | a**, Phase modulation as a function of the width *w* of the square unit cell made of crystalline silicon nanopillar under the 632 nm light illumination, the period of the unit cell is 260 nm. **b**, Transmittance as a function of the width *w* of the square unit cell made of crystalline silicon nanopillar

## S6. Reduction of FLOPs

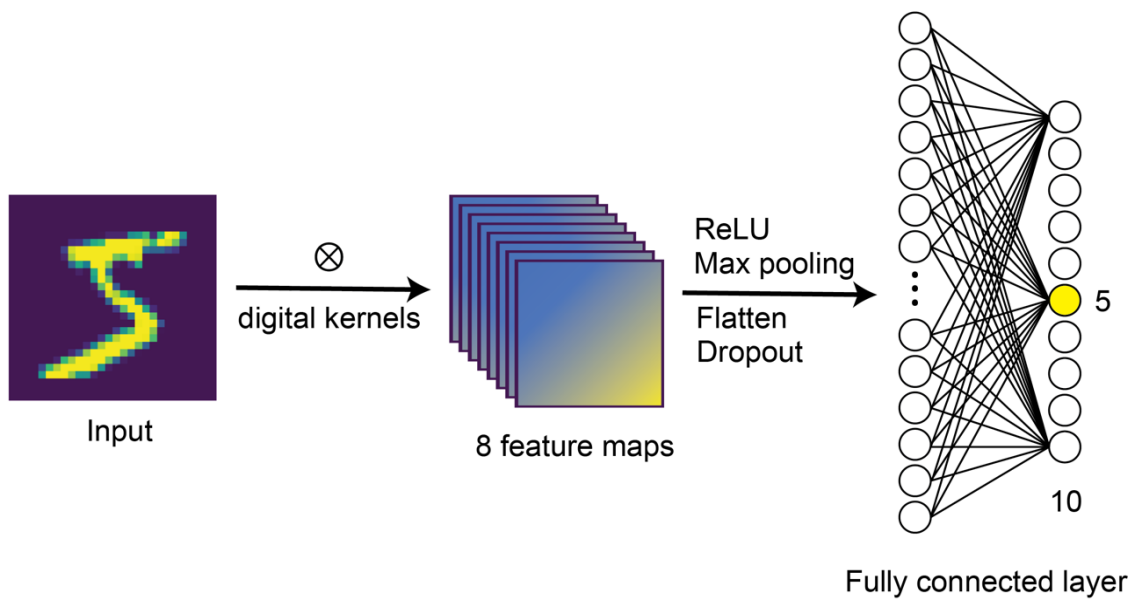

**Figure S4 |** The digital convolutional neural network corresponding to the ACNN features the same components. After convolution with digital kernels, we perform ReLU activation, max pooling, flattening, and dropout operations on the 8 feature maps. The resulting output is then fed into a fully connected layer for classification.

To calculate the reduction of FLOPs, a digital convolution neural network with the same components. is built, as shown in Fig. S4.

The FLOPs of a convolution layer can be calculated as,
$$FLOPs\_c = [(C_i \times k_w \times k_h) + (C_i \times k_w \times k_h - 1) + 1] \times C_o \times W \times H, \quad (S14)$$
where $C_i$ is the input channel number, $k_w$ and $k_h$ are the width and height of the kernel size respectively, $C_o$ is the output channel number, $W$ and $H$ are the width and height of the feature map respectively, $(C_i \times k_w \times k_h)$ is the FLOPs of multiplication in one convolution operation, $(C_i \times k_w \times k_h - 1)$ is the FLOPs of addition in one convolution operation, and + 1 is for the bias.

The FLOPs of one fully connected layer can be calculated as,
$$FLOPs\_f = [I + (I - 1) + 1] \times O. \quad (S15)$$
where $I$ is the number of input neurons, $O$ is the number of output neurons, and $[I + (I - 1) + 1]$ is the FLOPs of an output neuron including multiplication, addition, and bias.

|  | Network with 3 × 3 kernel size | Network with 9 × 9 kernel size |
| --- | --- | --- |
| FLOPs of convolution layer | 97344 | 518400 |
| FLOPs of fully-connected layer | 27040 | 16000 |

**Table S1** | FLOPs of the equivalent digital convolution neural networks

As shown in Table S1, the ACNN reduces the FLOPs compared with the fully digital neural network with the same components. The ACNN achieves a 78% reduction in FLOPs when using 3×3 convolution kernels in the digital neural network. When the kernel size is increased to 9×9, the ACNN can achieve an even greater reduction, reaching 97% in FLOPs.

## S7. Additional digital layers for complex tasks

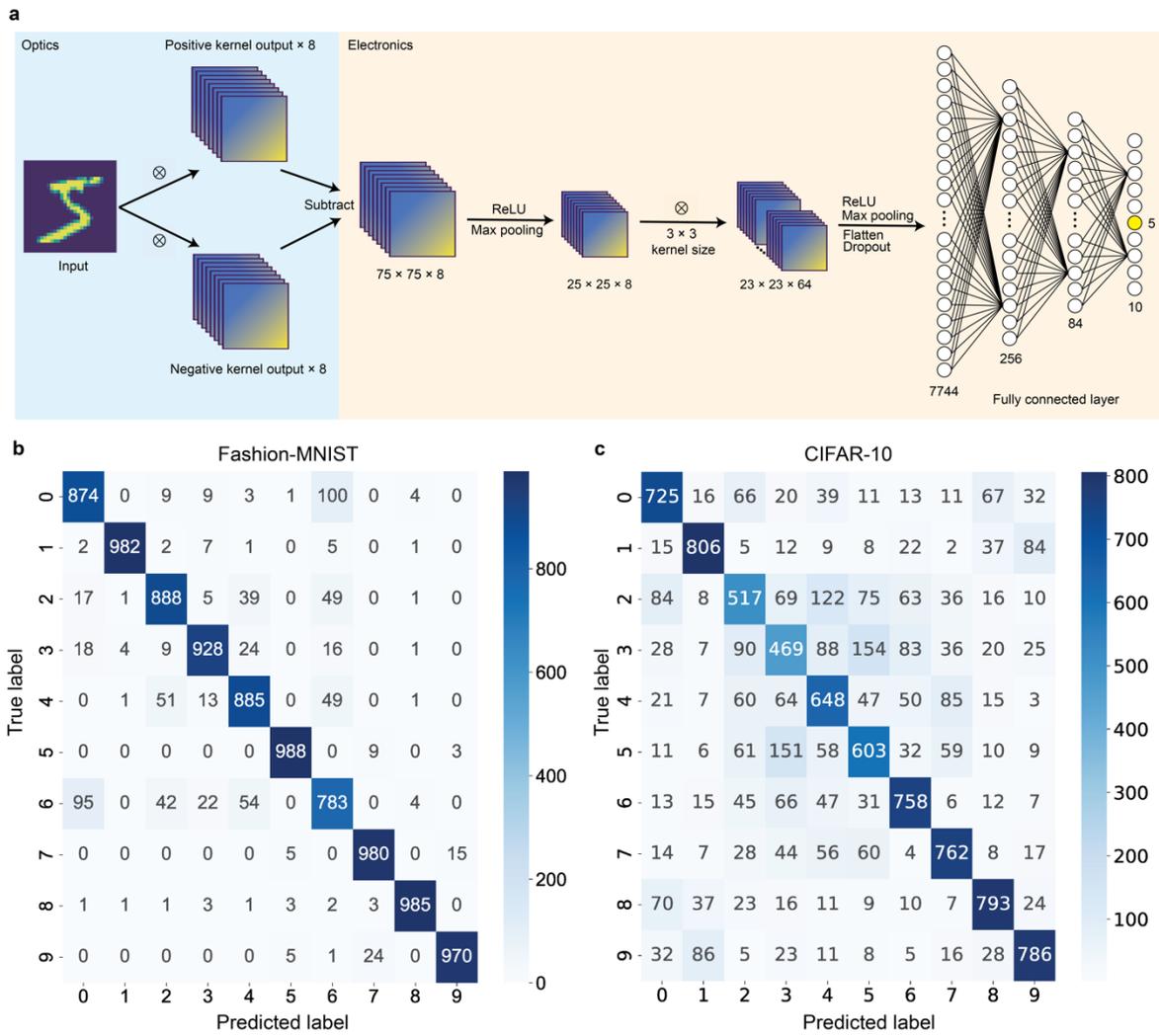

**Figure S5 | a**, Structure of the ACNN with additional digital layers. **b,** Simulated confusion matrix using validation images of the Fashion-MNIST dataset. **c,** Simulated confusion matrix using validation images of the CIFAR-10 dataset.

## S8. System robustness against quantization error

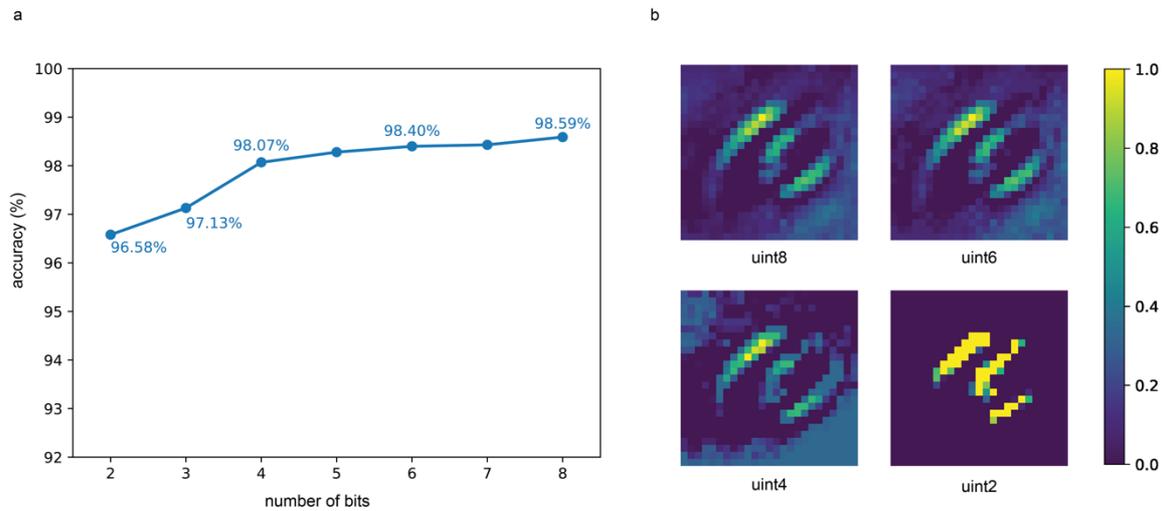

**Figure S6 | a**, Classification accuracy as a function of the bit number. **b**, Feature images in different bit numbers.

To test the robustness of the ACNN against quantization error, we quantify the data format of images taken by the camera to perform MNIST classification. The RAW image data format from the camera is an unsigned 8-bit integer (uint8). We quantize the RAW image to a lower-bit integer. After the re-training process of the digital fully connected layer, the classification accuracy as a function of the bit number is shown in Fig. S6, illustrating that the system can still effectively perform the classification task with reduced details.